\documentclass[english,journal=jctcce,manuscript=article,etalmode=truncate,maxauthors=0]{achemso}
\setkeys{acs}{etalmode=truncate,maxauthors=0}

\usepackage{longtable}
\usepackage{threeparttablex}
\usepackage{amsmath}             
\usepackage{amssymb}             
\usepackage{wasysym}             
\usepackage{color}               
\usepackage{setspace}            
\usepackage{graphicx}            
\usepackage{wrapfig}             
\usepackage[dvipsnames]{xcolor}  
\usepackage{soul}
\usepackage{array}
\usepackage{array,booktabs}      
\usepackage{bbm}
\usepackage{algorithm}
\usepackage{algpseudocode}
\usepackage{float}

\usepackage[font=footnotesize,labelfont=bf,labelsep=period,width=0.75\textwidth]{caption}   
\usepackage[font=footnotesize,labelfont=bf,labelsep=period]{subcaption}                     
\DeclareCaptionSubType*[arabic]{figure}                                                     

\usepackage[capitalize]{cleveref}
\crefname{figure}{Fig.}{Figs.}
\Crefname{figure}{Figure}{Figures}
\crefname{table}{Tab.}{Tabs.}
\Crefname{table}{Table}{Tables}
\crefname{equation}{Eq.}{Eqs.}
\Crefname{equation}{Equation}{Equations}
\crefname{section}{Sec.}{Secs.}
\Crefname{section}{Section}{Sections}

\title[]{General Symmetry-Based Potential Energy Surface Grid Reduction in Normal Coordinates}
\author{Can Liao}
\email{can.liao@phys.chem.ethz.ch}
\affiliation[ETH Zürich]
{ETH Zürich, Department of Chemistry and Applied Biosciences, Vladimir-Prelog-Weg 2, 8093 Zürich, Switzerland}
\author{Markus Reiher}
\email{mreiher@ethz.ch}
\affiliation[ETH Zürich]
{ETH Zürich, Department of Chemistry and Applied Biosciences, Vladimir-Prelog-Weg 2, 8093 Zürich, Switzerland}

\begin{document}

\begin{abstract}
The construction of a grid-based potential energy surface (PES) can be prohibitively expensive as the number of grid points grows exponentially with molecular size.
Molecular symmetry can reduce this cost by eliminating symmetry-equivalent points. 
We present an algebraic symmetry-based grid reduction method, ASyBGR, that is capable of handling non-Abelian and higher-order cyclic symmetry groups.
Non-coordinate-mixing operations are encoded into a vector space bit-string, where Gaussian elimination and closure can identify all symmetry-valid sign change patterns and the coordinate axes whose grids can be halved about the origin.
In the presence of degenerate subspaces, we optimize the coordinate basis to maximize reduction. 
The method was validated by comparing vibrational configuration interaction (VCI) energies calculated using the full and symmetry-reduced fourth-order HDMR PESs for molecules spanning a broad range of symmetries. 
The number of grid points required to construct the PES was reduced by as much as 81\%, while VCI energies deviated below $1$ cm$^{-1}$ for all test cases on average.
The relative root-mean-square deviation (RRMSD) between the full and symmetry-reduced potential energy and dipole moment surfaces were at most on the order of $10^{-3}$ and $10^{-2}$, respectively.
\end{abstract}

\section{Introduction}
Within the Born-Oppenheimer approximation, the potential energy surface (PES) is usually a prerequisite for nuclear dynamics and spectroscopic simulations.\cite{Bowman1986_202, Meyer1990_73, Carter1997_10458, Beck2000_1, Bowman2003_533, Christiansen2003_5773, Christiansen2004_2140, Christiansen2004_2149, Boese2005_863, Christiansen2005_194105, Rauhut2007_184109, Bowman2008_2145, Petit2013_7009, Garnier2016_204123, Baiardi2017_3764, Christiansen2019_MidasCpp, Larsson2019_204102, Fetherolf2021_074104, Bowman2022_Book, Glaser2023_9329}
Considerable effort has been devoted to the development of PES representations and parameterization methods.\cite{Henry1965_168, Henry1961_319, Jackle1996_7974, Li2001_7765, Rauhut2004_9313, Manzhos2006_084109, Rauhut2007_184109, Manzhos2008_224104, Lin2008_23, Sparta2009_8712, Sparta2010_3162, Xie2010_26, Csaszar2012_273, Sibaev2015_2200, Avila2015_044106, Ziegler2016_114114, Tan2018_6405, Boussaidi2020_7598}
A shared commonality between many of these methods is that, at some stage, the PES must be evaluated on a grid of molecular structures.
Depending on the desired accuracy, the required electronic structure calculations can be extremely expensive, while the number of grid points grows exponentially with the dimensionality of the PES.\\

Molecular symmetry is routinely exploited to reduce the computational cost of PES construction.\cite{Yagi2000_1005, Huang2002_8182, Wang2003_94, Bowman2003_533, Rauhut2004_9313, Wang2005_154303, Feller2009_154306, Sparta2010_3162, Pradhan2013_6925, Oschetzki2014_16426, Nikitin2016_114309, Ziegler2018_164110, Tan2018_6405, Christiansen2019_MidasCpp, Mitoli2023_3671, Seko2024_214302, Schneider2024_094102}
Within each set of symmetry-equivalent geometries, the PES needs to be evaluated at only one representative geometry.
The resulting energy can then be assigned to all other molecular structures in the set.
For vector and tensor properties, the individual components are not generally identical and must instead be transformed along side the molecular geometry by the same symmetry operations.\cite{Altmann1986_Book}\\

Many existing implementations of this reduction scheme are restricted to a small number of Abelian symmetry groups, whose symmetry operations map normal coordinates either onto itself or onto its negative.\cite{Bowman2003_533, Christiansen2019_MidasCpp}
This simplifies the implementation because the coordinate transformations can be determined directly from irreducible representations (irreps) listed in the symmetry group's character table. For degenerate irreps that appear in non-Abelian symmetry groups, the character table only provides the trace of the representation matrices and does not specify how each coordinate transforms under a symmetry operation.
Although symmetry operations within the same conjugacy class have the same character, they may act differently on individual coordinates within a degenerate subspace.
For example, one operation may flip the sign of one coordinate while leaving its degenerate counterpart unchanged, whereas another may mix the two coordinates.
This forces one to examine the action of each symmetry operation on the individual coordinates rather than inferring from the character table.
While cyclic symmetry groups are Abelian, some symmetry operations leave coordinates with a complex phase.\cite{Altmann1994_Book}
When these transformations are restricted to a real coordinate basis, they act through non-diagonal matrices within pairs of coordinates, appearing as coordinate mixing.
This mirrors the difficulties similar to those encountered for multidimensional irreps of non-Abelian symmetry groups.\\

To the best of our knowledge, the only reduction method that generalizes the symmetry-based grid reduction scheme to an arbitrary symmetry group is brute-force sampling of the PES using an inexpensive approximate symmetry-preserving electronic structure method.\cite{Ziegler2018_164110}
This reduction method identifies symmetry-equivalent geometries through shared diagnostic energies.
The original implementation\cite{Ziegler2018_164110} of this reduction method showed that the energy derived from diagonalizing the electronic core Hamiltonian suffices as the diagnostic energy.
In principle, two non-equivalent geometries may coincidentally carry the same diagnostic energy, leading to a false positive.
The authors of the reduction method claim\cite{Ziegler2018_164110} that this is extremely rare and such coincidences were not observed within their test cases.
Nevertheless, sampling every point on a PES can become very expensive as the total number of points grows exponentially with the number of coordinates.
Switching over to using the nuclear-nuclear potential energy as the diagnosis energy does alleviate some computational burden, but this comes with a higher risk for false positives.
\\

In practice, molecular geometries may deviate slightly from exact symmetry.
In the context of the aforementioned reduction method, imperfect geometries and numerical thresholds may cause some symmetry-equivalent grid points to be missed.
Sensitivity to numerical thresholds does not only affect the reduction scheme described above, but nearly all computational methods that pertain on molecular symmetry.\cite{Wang2003_94, Ziegler2018_164110, Knowles2022_161, Gunde2024_062503} 
This is generally difficult whenever discrete algebraic relations are inferred from continuous quantities on computer systems, requiring discrete classifications to be made from numerical data.
An enormous body of work has been dedicated to the seemingly simple task of assigning a symmetry group to a molecule and molecular eigenvectors to irreps.\cite{Budai1977_97, Zabrodsky1992_7843, Pilati1998_503, Ivanov1999_728, Largent2012_1637, Johansson2017_8, GyeviNagy2017_156, Knowles2022_161, Huynh2024_114, Gunde2024_062503, Nielsen2024_5740, Nielsen2025_11122}
A general robust symmetry-based grid reduction method must not only cover all symmetry groups but also be resilient against numerical errors.\\

Here, we present an algebraic symmetry-based grid reduction method that is capable of handling non-Abelian and cyclic symmetry groups.
Rather than inferring coordinate transformations from irrep characters, our scheme examines the action of each detected symmetry operation on every normal coordinate and retains those operations that act exclusively through coordinate sign changes. 
Because the choice of coordinate bases within degenerate subspaces determines which operations appear coordinate-mixing, we also introduce a procedure to optimize the coordinate basis to maximize reduction.\\

Depending on the symmetry identification algorithm, only a subset of the symmetry operations may be observed directly during symmetry group assignment,\cite{Budai1977_97, Pilati1998_503, Atkins2010_Book, Knowles2022_161} requiring the remaining operations to be recovered through closure. 
This step can be numerically delicate for imperfect geometries because the corresponding matrix representations are only approximate. 
Accumulated errors may cause valid operations to be omitted or duplicate operations to be retained. 
To address this problem, we utilize a discrete algebraic representation of the non-coordinate-mixing operations that eliminates redundant operations and recovers missing coordinate sign-change relations without further floating-point matrix comparisons or error accumulations. 

\section{Theory}
A normal coordinate system is employed throughout this work.
A general PES can be decomposed into a high-dimensional model representation (HDMR) expansion, where each term depends on a subset of coordinates.\cite{Carter1997_10458, Li2001_7765, Manzhos2006_084109}
\begin{equation}
    V(\mathbf{q}) = \sum\limits_i V^{(i)}(q_i) + \sum\limits_{i<j} V^{(ij)}(q_i,q_j) + \sum\limits_{i<j<k} V^{(ijk)}(q_i,q_j,q_k) + \hdots + V^{(\mathcal{M})}(\mathbf{q}).
\end{equation}
Each expansion term is constructed through inclusion-exclusion
\begin{equation}\label{eq:HDMRTerms}
    V^{(\mathcal{S})}(\mathbf{q}_\mathcal{S}) = \sum\limits_{\mathcal{T}\subseteq\mathcal{S}}(-1)^{|\mathcal{S}| - |\mathcal{T}|}V(\mathbf{q}_{\mathcal{T}}\oplus\mathbf{q}_{\mathcal{M}\setminus\mathcal{T}}^{(0)}),
\end{equation}
where $\mathcal{S}$ is the set of coordinates coupled by $V^{(\mathcal{S})}$.
Each summation term in \Cref{eq:HDMRTerms} adds contributions or removes over-counting from a subset of coordinates $\mathcal{T}\subseteq\mathcal{S}$.
This term is based on evaluating the PES at $\mathbf{q}_\mathcal{S}\oplus \mathbf{q}_\mathcal{\mathcal{M}\setminus\mathcal{T}}^{(0)}$, where $\mathbf{q}_\mathcal{S}$ is a coordinate vector in the coordinate subspace spanned by $\mathcal{S}$ and $\mathbf{q}_{\mathcal{M}\setminus\mathcal{T}}$ denotes the origin in the complementary subspace spanned by $\mathcal{T}$.
The HDMR representation allows for a well-defined hierarchical approximation of the PES through the truncation of the expansion.
The PES is materialized as a grid of values for each expansion term.
Typically, this representation is only tractable at low truncation orders as the number of grid points required for each term increases exponentially with the dimensionality of the term.
This representation is not restricted to only the PES, but to any function of nuclear positions, such as multipole moments.

\subsection{Symmetry-Based Grid Reduction Scheme Overview}
Molecular symmetry is described by the collection of symmetry operations that leave the molecular structure invariant.
This collection satisfies the mathematical properties of a \textit{group} and is therefore called the symmetry group $\mathcal{G}$.
Molecular symmetry groups are typically finite, with the exception of linear molecules.
The presented reduction method is only applicable to finite symmetry groups, so the linear symmetry groups $D_{\infty h}$ and $C_{\infty v}$ must be lowered to the finite $D_{2h}$ and $C_{2v}$ symmetry groups, respectively, since any $2\pi/n$ rotation is coordinate-mixing for $n>2$.
Two points are symmetry-equivalent if there exists one symmetry operation $g\in\mathcal{G}$ such that
\begin{equation}
    \mathbf{q}' = \mathbf{D}^g\mathbf{q}
\end{equation}
where $\mathbf{D}^g$ is the matrix representation of $g$ in this coordinate system.

Symmetry-equivalent points correspond to the same relative positioning of atoms in the molecule.
In the absence of an external potential, the two points must share the same molecular properties, up to some reorientation of tensor observables.
Therefore, scalar observables, such as energies, are invariant with respect to symmetry operations ($V(\mathbf{q}) = V(\mathbf{D}^g\mathbf{q})$).
Tensor observables, such as multipole moments, are not generally invariant.
Instead, they are equivariant and must be transformed alongside the molecular structure.\cite{Altmann1986_Book}
For example, the dipole moments $\boldsymbol{\mu}$ between symmetry-equivalent points are related by
\begin{equation}
    \boldsymbol{\mu}(\mathbf{D}^g\mathbf{q}) = \mathbf{R}^g\boldsymbol{\mu}(\mathbf{q})
\end{equation}
where $\mathbf{R}^g$ is the matrix representation of the symmetry operation $g$ in Cartesian coordinates.

\subsection{Bit-String Representation of Non-Coordinate-Mixing Symmetry Operations}
The crux of this method is to identify sets of symmetry-equivalent points, which we call symmetry orbits. 
Rather than working with the symmetry operations themselves, it is more convenient to work with a representation of their actions onto the coordinates. 
Since all coordinate-mixing symmetry operations are omitted, the ones remaining form a line-stabilizer group $\mathcal{G}'$ where each operation can be described by a series of coordinate sign changes. 
This can be encoded as a bit-string where $0$ indicates no change and $1$ indicates a sign change, and the length of the bit-string equals the number of coordinates.
The encoding map can be written as
\begin{equation}
\begin{split}
    &F: \mathcal{G}'\rightarrow\mathbb{B}\\
    &g \mapsto \mathbf{b} = b_0b_1\hdots b_{N-1}
\end{split}
\end{equation}
where $N$ is the number of coordinates and $\mathbb{B}$ is the set of bit strings of length $N$.
We denote the action represented by $\mathbf{b}$ onto a coordinate vector with ($\cdot$):
\begin{equation}
    \mathbf{b}\cdot\mathbf{Q} = 
    \begin{pmatrix}
        (-1)^{b_0}q_0 \\
        \vdots \\
        (-1)^{b_i}q_{i} \\
        \vdots \\
        (-1)^{b_{N-1}}q_{N-1}
    \end{pmatrix}
\end{equation}
This map is neither injective nor surjective.
Multiple symmetry operations can be mapped to the same bit-string, providing a convenient elimination of redundant symmetry operations. 
At the limit where $F$ is surjective, the maximum grid reduction is achieved, that is, the symmetry orbit of a point includes every possible sign combination of its coordinates.\\

The encoding map serendipitously maps the line-stabilizer $\mathcal{G}'$ onto a more convenient algebra.
The set of bit-strings $\mathbb{B}$ is an $N$-dimensional vector space over $GF(2)$, the finite field of two bit-states.
Elements within a mathematical field interact via two closed binary operations called a \textit{multiplication} or an \textit{addition}, where the multiplication is distributive over the addition.
In $GF(2)$, the logical $\operatorname{AND}$ and $\operatorname{XOR}$ operators act as the multiplication ($\land$) and addition ($\oplus$), respectively.\\

The vector space $\mathbb{B}$ is formally the $N$-fold Cartesian product of $GF(2)$.
Addition defined on $\mathbb{B}$ is inherited from $GF(2)$ as the bitwise $\operatorname{XOR}$.
The composition of symmetry operations manifests as the bitwise $\operatorname{XOR}$ operation, guaranteeing commutativity even if $\mathcal{G}'$ is non-Abelian.
Scalar multiplication of an element $a\in GF(2)$ onto a bit-string is defined as
\begin{equation}
    a\land\mathbf{b} = (a\land b_0)(a\land b_1)\hdots(a\land b_{N-1})
\end{equation}
Any bit-string can be written as a linear combination of basis strings $\mathbf{e}_i$
\begin{equation}
    \mathbf{b} = \bigoplus\limits_{i=0}^{N-1} a_i\land\mathbf{e}_i
\end{equation}
where $a_i\in GF(2)$.
The fact that $\mathcal{G}'$ is a group causes its encoded image $F(\mathcal{G}')$ to be a vector subspace of $\mathbb{B}$.\\

\subsection{Finding Symmetry Orbits}
Suppose the nuclear configuration space is spanned by the set of normal coordinates $\mathcal{S}$.
We begin by identifying the symmetry group.
The set of diagnostic operations collected in this process may definitively determine the symmetry group, but may only be a subset of the full group.
In theory, the entire symmetry group can be recovered from the closure of the set of diagnostic operations.
Because of possible numerical complications, we proceed without assuming that closure yields the exact group $\mathcal{G}$, but rather some set of symmetry operations $\mathcal{H}$.
Coordinate-mixing symmetry operations are still removed from $\mathcal{H}$ to yield $\mathcal{H}'$.\\

Unlike $\mathcal{G}'$, $\mathcal{H}'$ is not necessarily a group, hence, $F(\mathcal{H}')$ may not be a vector subspace of $\mathbb{B}$ before closure.
The closure of $F(\mathcal{H}')$, denoted as $\mathbb{H}$, has a maximum dimension of $|\mathcal{S}|$.
At the limit where $\mathrm{dim}(\mathbb{H}) = |\mathcal{S}|$, $\mathbb{H}$ becomes $\mathbb{B}$, corresponding to the maximum symmetry reduction case.
Undergoing closure with $F(\mathcal{H}')$ is equivalent to constructing a linearly independent basis from $F(\mathcal{H}')$ and finding the span of that basis.
The size of $F(\mathcal{H}')$ may be larger than the dimension of $\mathbb{H}$, or even $|\mathcal{S}|$.
In that case, while $F(\mathcal{H}')$ itself can be used as a basis, it is not ideal since linear dependencies can lead to unnecessary redundancies during closure.\\

A linearly independent basis can be extracted through Gaussian elimination of a matrix with rows constructed from bit strings in $F(\mathcal{H}')$.
\begin{equation}
    \begin{pmatrix}
        \mathbf{b}_1 \\
        \mathbf{b}_2 \\
        \vdots \\
        \mathbf{b}_{|F(\mathcal{H}')|}
    \end{pmatrix}
    \xrightarrow{\text{Gaussian elimination}}
    \begin{pmatrix}
        \mathbf{e}_1 \\
        \mathbf{e}_2 \\
        \vdots \\
        \mathbf{e}_{|\mathbb{H}|} \\
        \mathbf{0} \\
        \vdots \\
        \mathbf{0}
    \end{pmatrix}
\end{equation}
The remaining non-zero rows in the row-echelon matrix form a linearly independent basis.
Closure is performed by taking all possible linear combinations of the basis strings.
\begin{equation}
    \mathbb{H} = \Bigg\{\bigoplus\limits_{i = 1}^{|\mathbb{H}|} a_i\land\mathbf{e}_i\Bigg|\forall a_i\in GF(2)\Bigg\}
\end{equation}
For a given point $\mathbf{Q}_\mathcal{S}$, its symmetry orbit
\begin{equation}
    \Omega(\mathbf{q}_\mathcal{S}) = \{\mathbf{h}\cdot\mathbf{q}_\mathcal{S}|\forall\mathbf{h}\in\mathbb{H}\}
\end{equation}
is obtained by applying the symmetry action represented by bit strings in $\mathbb{H}$.
Hence, the size of the orbit is equal to the size of $|\mathbb{H}|$. 
Since $GF(2)$ only has two elements,
\begin{equation}\label{eq:OrbitSize}
    |\Omega(\mathbf{q}_\mathcal{S})| = |\mathbb{H}| = 2^{\mathrm{dim}(\mathbb{H})}
\end{equation}
In practice, we recommend undergoing closure within the Cartesian matrix representation as well as the bit-string representation.
Because two coordinate-mixing operations can yield a non-coordinate-mixing operation, solely relying on closure in the bit-string representation can leave behind valid sign-change patterns.
Closure in the bit-string representation acts as a backup to recover as many sign-change patterns as possible when closure fails in the Cartesian matrix representation.

\subsection{Grid Point Reduction Scheme}
The row-echelon matrix provides a way to identify the coordinates whose grids can be halved about the origin while ensuring that all omitted grid points can be recovered through symmetry.
Suppose we desire to compute only the negative half-grid along as many coordinates as possible.
The pivot columns identify coordinates whose sign-change pattern can be chosen independently when selecting among the available symmetry actions.
In contrast, the sign changes of the non-pivot coordinates are determined by the chosen pattern of the pivot coordinates.
For example, let column $i$ be a pivot column and $\mathbf{b}\in\mathbb{H}$.
Suppose that
\begin{equation}
    \mathbf{b}=\mathbf{l}b_i\mathbf{r}
\end{equation}
where $\mathbf{l}$ and $\mathbf{r}$ contain the bits to the left and right of position $i$, respectively.
The pivot bit $b_i$ can change without the bit-string leaving $\mathbb{H}$, regardless of other pivot bits.
Changing $b_i$ may change the non-pivot bits contained in $\mathbf l$ and $\mathbf r$ as those bits depend on the complete pivot-bit pattern.
Consequently, there always exists $\mathbf{b}\in\mathbb{H}$ such that $\mathbf{b}$ maps $\mathbf{q}_{\mathcal{S}}$ to a point where all pivot coordinates are negative.
This is not true for non-pivot coordinates as they are not freely chosen.
Therefore, only the negative half-grids of the pivot coordinates need to be computed.
For a grid where the number of non-origin grid points along each coordinate $i\in\mathcal{S}$ is $d_i$, the number of grid points in the full multidimensional grid spanned by $\mathcal{S}$
\begin{equation}
    N_{\mathrm{total}}^\mathcal{S} = \prod\limits_{i\in\mathcal{S}}d_i
\end{equation}
is reduced to
\begin{equation}\label{eq:Nreduced1}
    N_{\mathrm{reduced}}^\mathcal{S} = \frac{N_{\mathrm{total}}^\mathcal{S}}{|\mathbb{H}|}
\end{equation}

Recall that tensor observables are equivariant, meaning that the property tensor must be transformed alongside the coordinate vector.
Reorienting the property tensor requires mapping the bit-strings back to symmetry operations.
This will trivial if the symmetry group is exactly recovered from closure of the diagnostic set.
Even though the encoding map is not injective, it suffices to use one symmetry operation from the bit-string pre-image to transform the property tensor.
If the symmetry group is not exactly recovered and only an incomplete set of symmetry operations have been accepted, the series of $\operatorname{XOR}$ operations done during Gaussian elimination and $F(\mathcal{H}')$ closure must be tracked to recreate the symmetry operations responsible for bit strings obtained during $F(\mathcal{H}')$ closure.
This is because these symmetry operations are initially rejected based on numerical thresholding even though they should be accepted based on the property of group closure under composition.

\subsection{Example: Five-Mode Grid Reduction in $D_{6h}$}
To illustrate the execution of this method, we provide an example with $D_{6h}$ symmetry.
Consider the grid representation of the potential term involving $\mathcal{S}=(i,j,k,l,m)$, whose irreps are $B_{2g}$, $A_{2u}$, $B_{1u}$, $E_{1g}$, and $E_{1g}$, respectively.
Here, $l$ and $m$ form a degenerate subspace.
To provide a tractable example that can be worked through by hand and demonstrate the robustness of this reduction scheme, suppose the closure of the diagnostic set yields
\begin{equation}
    \mathcal{H} = \{E, C_6, \sigma_h, C_2, C_2', \sigma_v\}
\end{equation}
where $E$ is the identity element, $C_6$ is a six-fold rotation, $C_2$ and $C_2'$ are two-fold rotations, and $\sigma_h$ and $\sigma_v$ are mirror planes.
The rotations $C_2$ and $C_6$ are coaxial while $C_2'$ is orthogonal to them.
$\sigma_h$ is coplanar with the molecular plane.
$\sigma_v$ is orthogonal to the molecular plane.
$C_6$ is omitted because it mixes the two degenerate $E_{1g}$ coordinates.
\begin{equation}
    \mathcal{H}' = \{E, \sigma_h, C_2, C_2', \sigma_v\}.
\end{equation}

The encoded image of $\mathcal{H}'$ is
\begin{equation}
    F(\mathcal{H}') = \{0000, 11011, 10111, 11001, 10101\}.
\end{equation}
The non-zero bits are placed into a matrix and cast into row-echelon form.
\begin{equation}
    \begin{pmatrix}
        1 & 1 & 0 & 1 & 1\\
        1 & 0 & 1 & 1 & 1\\
        1 & 1 & 0 & 0 & 1\\
        1 & 0 & 1 & 0 & 1
    \end{pmatrix}
    \xrightarrow{\text{Gaussian elimination}}
    \begin{pmatrix}
        1 & 1 & 0 & 1 & 1\\
        0 & 1 & 1 & 0 & 0\\
        0 & 0 & 0 & 1 & 0\\
        0 & 0 & 0 & 0 & 0
    \end{pmatrix}.
\end{equation}
The pivot columns correspond to coordinates $i$, $j$, and $l$, indicating that only the negative half of the grid along these coordinates needs to be computed.
Each bit string $\mathbf{h}\in\mathbb{H}$ can be written as
\begin{equation}
    \mathbf{h} = (a_0\land 11011) \oplus (a_1\land 01100) \oplus (a_2\land 00010)
\end{equation}
for some $a_0, a_1, a_2\in GF(2)$.
Applying all combinations of $(a_0, a_1, a_2)$ yields the entirety of $\mathbb{H}$:
\begin{equation}
    \mathbb{H} = \{00000, 11011, 01100, 00010, 10111, 11001, 01110, 10101\}.
\end{equation}
Applying all $\mathbf{h}\in\mathbb{H}$ onto $\mathbf{q}_\mathcal{S}$ yields its orbit.
\begin{equation}
\begin{split}
    &\Omega(\mathbf{q}_\mathcal{S}) \\
    &=
    \left\{
    \begin{matrix}
    \begin{pmatrix}
        q_i \\ q_j \\ q_k \\ q_l \\q_m
    \end{pmatrix},
    &
    \begin{pmatrix}
        -q_i \\ -q_j \\ q_k \\ -q_l \\-q_m
    \end{pmatrix},
    &
    \begin{pmatrix}
        q_i \\ -q_j \\ -q_k \\ q_l \\q_m
    \end{pmatrix},
    &
    \begin{pmatrix}
        q_i \\ q_j \\ q_k \\ -q_l \\q_m
    \end{pmatrix},
    &
    \begin{pmatrix}
        -q_i \\ q_j \\ -q_k \\ -q_l \\-q_m
    \end{pmatrix},
    &
    \begin{pmatrix}
        -q_i \\ -q_j \\ q_k \\ q_l \\-q_m
    \end{pmatrix},
    &
    \begin{pmatrix}
        q_i \\ -q_j \\ -q_k \\ -q_l \\q_m
    \end{pmatrix},
    &
    \begin{pmatrix}
        -q_i \\ q_j \\ -q_k \\ q_l \\-q_m
    \end{pmatrix}
    \end{matrix}
    \right\}.
\end{split}
\end{equation}

\subsection{Optimization of Degenerate Subspace Bases for Maximum Reduction}
Coordinate-mixing symmetry operations are represented by non-diagonal matrices in the coordinate representation.
Whether a symmetry operation is coordinate-mixing depends on the chosen coordinate basis.
Within a degenerate subspace, rotating the normal coordinate basis yields another valid normal coordinate basis while changing the matrix representation of the group.
Optimizing the degenerate subspace basis to maximize reduction does not necessarily mean minimizing coordinate-mixing operations since different operations may be represented by the same matrix.
Instead, the optimal basis for a degenerate subspace maximizes the number of unique diagonal matrix representations.
An additional constraint must be considered in the presence of multiple degenerate subspaces.
One must keep in mind that the set of symmetry operations that are non-coordinate-mixing in the optimized basis may differ for every subspace.
When both subspaces are considered simultaneously, only symmetry operations that are non-coordinate-mixing for both are kept.
Consequently, the optimization must also maximize the intersection of non-coordinate-mixing symmetry operations corresponding to the chosen basis for each subspace.\\

If there exists a basis in which a symmetry operation leads only to coordinate sign changes, then all eigenvalues of its matrix representation must be $\pm 1$.
These matrices, called involutions, are self-inverting, and must be symmetric and orthogonal.\cite{Bernstein2018_Book}
This comes with the salient property that two symmetric matrices are simultaneously orthogonally diagonalizable if and only if they commute.
Candidate bases for each degenerate subspace can be generated by simultaneously diagonalizing maximal subsets of commuting involutions in the subspace representations.\\

Let $\mathbb{X}(\mathcal{U})$ be the irreducible invariant subspace in which the set of degenerate normal coordinates $\mathcal{U}$ resides.
For completeness, we consider a non-degenerate coordinate as a trivially degenerate set of one coordinate.
However, we continue to only refer to multidimensional irreducible invariant subspaces as ``degenerate subspaces''.
A set of candidate bases can be generated for each degenerate subspace $\mathbb{X}(\mathcal{D})$, where $\mathcal{D}$ is a non-trivial set of degenerate coordinates.
After selecting a candidate basis $\mathbf{B}_{\mathbb{X}(\mathcal{D})}$ for each degenerate subspace $\mathbb{X}(\mathcal{D})$, the global candidate basis $\mathbf{B}$ can be constructed as a direct sum of bases from each irreducible invariant subspace,
\begin{equation}
    \mathbf{B} = \bigoplus_{\mathcal{U}} \mathbf{B}_{\mathbb{X}(\mathcal{U})}
\end{equation}
The number of global candidate bases scales roughly exponentially with the number of degenerate subspaces.
Calculating symmetry orbit sizes using \Cref{eq:OrbitSize} becomes too costly.
Alternatively, we turn to group theory to find a more efficient method to compute the symmetry orbit size.\\

For each degenerate subspace $\mathbb{X}(\mathcal{D})$, we find all $g\in\mathcal{G}$ such that $g$ produces a sign change on each basis vector in $\mathbf{B}_{\mathbb{X}(\mathcal{D})}$.
This set is the line-stabilizer group $\mathcal{L}_{\mathbb{X}(\mathcal{D})}$ with respect to $\mathbf{B}_{\mathbb{X}(\mathcal{D})}$.
The global line-stabilizer group $\mathcal{L}$ is formed by intersecting $\mathcal{L}_{\mathbb{X}(\mathcal{U})}$ across all irreducible invariant subspaces $\mathbb{X}(\mathcal{U})$,
\begin{equation}
    \mathcal{L} = \bigcap_{\mathcal{U}}\mathcal{L}_{\mathbb{X}(\mathcal{U})}
\end{equation}
The orbit--stabilizer theorem can be used to calculate the symmetry orbit size,\cite{Goodman2014_Book}
\begin{equation}
    |\mathbb{H}| = \frac{|\mathcal{L}|}{|\mathcal{K}|}
\end{equation}
where $\mathcal{K}$ is the stabilizer group with respect to the global candidate basis.
This is simply
\begin{equation}
    \mathcal{K} = \bigcap_{i\in\mathcal{M}}\mathcal{K}_{i}
\end{equation}
where $\mathcal{K}_{i}$ is the stabilizer group with respect to coordinate $i$, that is, the set of $g\in\mathcal{G}$ that leave the coordinate vector $i$ invariant.
Line-stabilizer groups are indexed over irreducible invariant subspaces since they are shared between all basis vectors within an irreducible invariant subspace basis.
Stabilizer groups are indexed over individual coordinates since different basis vectors within a basis may experience different sign changes.\\

The optimal basis may differ for each HDMR truncation order or choice in coupled coordinates.
The number of grid points after reduction $N_{reduced}^\mathcal{S}$ must be calculated for each $\mathcal{S}\subseteq\mathcal{M}$ with a size equal to the truncation order $N_T$.
The total number of grid points after reduction is calculated from summing $N_{reduced}^\mathcal{S}$ across the combinatorial collection of $\mathcal{S}$,
\begin{equation}\label{eq:Nreduced2}
    N_{reduced} = \sum\limits_{\substack{\mathcal{S}\subseteq\mathcal{M} \\ |\mathcal{S}|=N_{T}}}N^{\mathcal{S}}_{reduced}
\end{equation}
Exacerbated by the pseudo-exponential scaling number of global candidate bases that must be tested, the optimization can quickly become unaffordable if done naively.\\

The $O(|\mathcal{M}|\;\mathrm{choose}\;N_T)$ time complexity for a scoring single global candidate basis can be reduced by an iterative algorithm that takes advantage of two observations: the reduction factor for $\mathcal{S}$ only depends on $\mathcal{L}_\mathcal{S}$ and $\mathcal{K}_\mathcal{S}$, and two different coordinate combinations may share a stabilizer pair.
These observations allow for the factorization of \Cref{eq:Nreduced2} into
\begin{equation}
    N_{reduced} = \sum\limits_{\mathcal{L},\mathcal{K}}\frac{|\mathcal{K}|}{|\mathcal{L}|}A_{|\mathcal{M}|,N_T}(\mathcal{L}, \mathcal{K})
\end{equation}
and
\begin{equation}\label{eq:InductiveInvariant}
    A_{m,n}(\mathcal{L}, \mathcal{K}) = \sum\limits_{\substack{\mathcal{S}\subseteq\mathcal{M}'(m),
    \; |\mathcal{S}| = n\\ \mathcal{L}_\mathcal{S} = \mathcal{L},\;\mathcal{K}_\mathcal{S} = \mathcal{K}}} \prod\limits_{i\in\mathcal{S}}d_i 
\end{equation}
where $\mathcal{M}'(m) = \{1,\hdots,m\}$. 
$A_{m,n}(\mathcal{L}, \mathcal{K})$ is the total number of grid points needed before reduction for all coordinate combinations of size $n$ that exhibit $\mathcal{L}$ and $\mathcal{K}$ as their line-stabilizer and stabilizer groups, respectively.
The algorithm iterates over $m$ until $m=|\mathcal{M}|$.
At each iteration, $A_{m,n}(\mathcal{L}, \mathcal{K})$ is computed for every triple $(n, \mathcal{L}, \mathcal{K})$ using $A_{m-1,n}(\mathcal{L}, \mathcal{K})$, where $0\leq n\leq N_T$.\\

The algorithm is initialized with
\begin{equation}
    A_{0,0}(\mathcal{G},\mathcal{G}) = 1
\end{equation}
with all other entries set to zero. 
At the beginning of each iteration $m$, $A_{m,n}(\mathcal L,\mathcal K)$ is initialized to zero.
Each subset of $\mathcal M'(m)$ at this iteration either excludes or includes the new coordinate $m$.
Subsets that exclude $m$ retain the same stabilizer pair, giving the update
\begin{equation}\label{eq:Update1}
    A_{m,n}(\mathcal L,\mathcal K)\leftarrow A_{m,n}(\mathcal L,\mathcal K) + A_{m-1,n}(\mathcal L,\mathcal K).
\end{equation}
Subsets that include $m$ acquire an additional factor of $d_m$, while their line-stabilizer and stabilizer groups are updated by intersection
\begin{equation}\label{eq:Update2}
    A_{m,n+1}
    (\mathcal L\cap\mathcal L_{\mathbb{X}(\mathcal{U})},
     \mathcal K\cap\mathcal K_{m})
    \leftarrow
    A_{m,n+1}
    (\mathcal L\cap\mathcal L_{\mathbb{X}(\mathcal{U})},
     \mathcal K\cap\mathcal K_{m}) + d_m A_{m-1,n}(\mathcal L,\mathcal K).
\end{equation}
where $m\in\mathcal{U}$.
This follows from the fact that adding a new coordinate introduces a new axis, which creates new grid points.
Although the sum in \Cref{eq:InductiveInvariant} is written over all $(\mathcal{L},\mathcal{K})$ pairs, these pairs do not need to be enumerated explicitly. 
Instead, the relevant $(\mathcal{L},\mathcal{K})$ pairs arise naturally from the intersections performed in \Cref{eq:Update2}.\\

The scoring algorithm is essentially a loop over accessible $(\mathcal{L},\mathcal{K})$ pairs, within a loop ranging from $1$ to $N_T$ and a loop over the set of normal coordinates. 
The time complexity for such a nested loop is $O(CN_T|\mathcal{M}|)$, where $C$ is the number of accessible $(\mathcal{L},\mathcal{K})$ pairs.
In the highly unlikely worst case scenario where every mode combination exhibits an unique $(\mathcal{L},\mathcal{K})$ pair, 
\begin{equation}
    \max(C) = \binom{|\mathcal{M}|}{N_T}
\end{equation}
With the reasonable assumption of a small $C$, the scoring algorithm scales linearly with $|\mathcal{M}|$ and $N_T$.\\

An honest complexity analysis must account for the need to score every global candidate basis.
The number of global candidate bases $B$ scales pseudo-exponentially with the number of degenerate subspaces because each degenerate subspace may have a different number of local candidate bases.
\begin{equation}
    B \approx L^D
\end{equation}
The number of degenerate subspaces $D$ can be generously capped by
\begin{equation}
    D =\Bigg\lfloor{\frac{|\mathcal{M}|-1}{r}}\Bigg\rfloor
\end{equation}
where $r=2$ for all symmetry groups except for $I$ and $I_h$ which has $r=3$.\cite{Altmann1994_Book}
The number of local candidate bases $L$ is bounded by the number of maximal Abelian involution subgroups in the subspace representation.
The highest number of maximal involution subgroups within a chemically relevant finite symmetry group is five, held by $I$ and $I_h$.\cite{Altmann1994_Book}
Therefore, the worst-case total time complexity of the coordinate basis optimization is approximately $O(5^{|\mathcal{M}|/2}|\mathcal{M}|N_T)$. \\

This is a very loose upper bound on the complexity of the optimization procedure.
Icosahedral molecules are extremely rare in chemistry so $L = 5$ is unlikely.
More common octahedral and $T_d$ symmetries have $L = 4$.
Theoretically, $L$ grows boundlessly with the rotation order of dihedral and $C_{nv}$ symmetry groups, but molecular systems typically do not possess $n > 6$. This leaves a realistic $L = 5$ for $n = 5$ and $L = 3$ for $n = 6$.\cite{Altmann1994_Book}
So far, we have only looked at $L$ for faithful representations.
In most cases, subspace representations are not faithful and can only decrease $L$.
Additionally, the exponential scaling with respect to $|\mathcal{M}|$ is also exaggerated as the number of degenerate subspaces is often much lower than the theoretical maximum.
With that being said, we can rationalize that the number of global candidate bases is small for most chemically relevant systems.
Unfortunately, we are unable to place a tighter bound on the time complexity with respect to the number of normal coordinates.
Nevertheless, with the number of normal coordinates kept constant, the optimization procedure scales linearly with $N_T$.\\

This intensive optimization is not necessary for linear symmetry groups.
A shortcut can be formulated by recognizing that all degenerate coordinates transform as the same irrep as either $(x,y)$ or the $(R_x,R_y)$, assuming the molecule is aligned with the $z$-axis.
The coordinate basis is optimized by using the $\sigma_{yz}$ (or $\sigma_{xz}$) eigenbasis for each degenerate subspace.\\

The efficient coordinate basis optimization procedure hinges on the orbit-stabilizer theorem to derive a simple analytical formula for the symmetry orbit size.\cite{Goodman2014_Book}
The orbit-stabilizer theorem requires that the set of actions, in this case, the signed-stabilizers, form a group.
If closure fails to recover the full symmetry group, the orbit-stabilizer theorem cannot be used and this optimization cannot be done efficiently.
Nevertheless, the reduction algorithm can be applied without optimizing the coordinate basis in most cases.

\section{Results and Discussion}
In this section, we demonstrate the validity and scaling of our algebraic symmetry-based grid reduction (ASyBGR) method.
ASyBGR was implemented within \textsc{Colibri}, our software for vibrational spectroscopic simulations.\cite{Glaser2023_Colibri}
We validated ASyBGR by examining the difference in VCI energies resulting from a symmetry-reduced PES and a full PES.
The VCI configuration basis was constructed from VSCF-optimized modals.
These modals are constructed from a discrete variable representation (DVR) basis induced from particle-in-a-box eigenfuncions.\cite{Colbert1992_1982, Light2000_263, Glaser2023_9329}
Details on the VCI calculations are found in the Supporting Information.\\

Demonstrating the validity of the method does not require agreement with experiment, but rather showing that results remain the same when symmetry reduction is used.
Computational efficiency is more important, as it enables validation on all coordinates and more symmetries.
Hence, it sufficed to construct the PES using the fast semi-empirical PM6 method.\cite{Stewart2007_1173, Husch2018_e25799, Bosia2023_054118}
The PES was constructed as a fourth-order HDMR expansion using all normal coordinates with seven points (including the origin) along each coordinate.
Both the reduced and full PESs were constructed with optimized normal coordinates.
The non-linear molecular structures optimized with PM6 displayed nearly ideal symmetries, with near-zero continuous symmetry operation measures.\cite{Nielsen2024_5740, Nielsen2025_11122}
C$_2$H$_2$ and HCN optimized to nearly perfect linear structure using PM6, but the continuous symmetry operation measures were not computed as it is not defined for linear symmetries.
All symmetry operations within their respective symmetry groups were successfully recovered.\\

We verify that our reduction method does not change  the physics of the calculation by showing that the maximum (MAD) and average absolute deviation (AAD) of the VCI energies are minimal.
The PES was constructed along all normal coordinates for an unbiased assessment of the effectiveness of the reduction scheme.
In all test cases shown in \Cref{tab:data1}, VCI energies computed using the symmetry-reduced grid deviated less than $1$ cm$^{-1}$ on average.
Only C$_6$H$_6$ exhibited a maximum deviation of greater than $1$ cm$^{-1}$.
This discrepancy can be traced back to a small number of fourth-order PES points corresponding to highly-compressed geometries.
The PES in this region is very steep so energy differences due to imperfect geometries are amplified.
Nevertheless, the maximum deviation between VCI energies calculated using a reduced PES and a full PES was merely $1.655$ cm$^{-1}$.

\begin{table}[H]
    \caption{Performance of the symmetry-based grid reduction scheme on the PES. The percentage of eliminated grid points (\% Reduced) and the maximum (MAD) and average absolute deviations (AAD) of the resulting VCI energies from those obtained with the full PES shown. Percentages in parentheses indicate the reduction achieved without coordinate basis optimization.}
    \label{tab:data1}
\begin{tabular}{lccccc}
\toprule
& \textbf{Symmetry}
& \textbf{\% Reduced}
& \textbf{MAD ($\mathrm{cm}^{-1}$)}
& \textbf{AAD ($\mathrm{cm}^{-1}$)} \\
\midrule
1,2-dibromo-1,2-dichloroethane & $C_i$ & 49\% & 0.132 & 0.005 \\
H$_2$O & $C_{2v}$ & 43\% & 0.002 & 0.000 \\
trans-C$_2$H$_2$Cl$_2$ & $C_{2h}$ & 66\% & 0.005 & 0.001 \\
CFH$_3$ & $C_{3v}$ & 43\% (0\%) & 0.495 & 0.019 \\
B$_2$H$_4$ & $D_{2d}$ & 71\% (50\%) & 0.757 & 0.201 \\
C$_2$H$_4$ & $D_{2h}$ & 80\% & 0.012 & 0.001 \\
C$_2$H$_6$ & $D_{3d}$ & 69\% & 0.052 & 0.009 \\
PCl$_5$ & $D_{3h}$ & 67\% (55\%) & 0.014 & 0.005 \\
S$_8$ & $D_{4d}$ & 71\% (52\%) & 0.053 & 0.007 \\
CuCl$_4^{2-}$ & $D_{4h}$ & 81\% (80\%) & 0.001 & 0.000 \\
C$_5$H$_5^-$ & $D_{5h}$ & 66\% (41\%) & 0.786 & 0.078 \\
C$_6$H$_6$ & $D_{6h}$ & 80\% & 1.655 & 0.319 \\
HCN & $C_{\infty v}$ & 67\% & 0.004 & 0.001 \\
C$_2$H$_2$ & $D_{\infty h}$ & 81\% & 0.010 & 0.003 \\
CH$_4$ & $T_{d}$ & 72\% (0\%) & 0.715 & 0.114 \\
SF$_6$ & $O_{h}$ & 81\% (49\%) & 0.063 & 0.010 \\
\bottomrule
\end{tabular}
\end{table}

The reduction scheme substantially reduced the number of grid points required to construct the PES.
For high-symmetry cases, the number of required grid points was reduced by as much as $81\%$. 
Although smaller reductions were expected for low-symmetry systems, the corresponding test cases still achieved reductions exceeding $40\%$.
Coordinate optimization played a crucial role in achieving such large reductions.
Without coordinate optimization, the grid of CH$_4$ and CFH$_3$ was irreducible.
These species exhibited only degenerate normal modes aside from the totally symmetric modes.
None of the degenerate normal coordinates happened to be eigenvectors of non-trivial symmetry operations in their respective symmetry groups. 
Consequently, all operations were labeled as coordinate-mixing and omitted.
Less drastic cases of this can be seen across the dihedral symmetry cases and $O_h$, where coordinate optimization contributed to a large percentage of the reduction.
This does not serve as a prescription on when coordinate optimization is needed.
Non-optimized normal coordinates may differ depending on the underlying electronic structure method, molecular orientation, or even eigensolver.
The only insight from this analysis is that coordinate optimization should be done, if possible, in the presence of degenerate subspaces.\\

\begin{table}[H]
\caption{The relative root-mean-square deviation (RRMSD) between the full and symmetry-reduced potential energy and dipole moment surfaces for each test case.}
\label{tab:data2}
\begin{tabular}{lcc}
\toprule
& \textbf{PES RRMSD}
& \textbf{Dipole RRMSD} \\
\midrule
1,2-dibromo-1,2-dichloroethane & $6.7\times10^{-4}$ & $1.8\times10^{-2}$ \\
H$_2$O & $8.5\times10^{-7}$ & $2.1\times10^{-6}$ \\
trans-C$_2$H$_2$Cl$_2$ & $4.7\times10^{-6}$ & $3.0\times10^{-5}$ \\
CFH$_3$ & $1.2\times10^{-4}$ & $6.2\times10^{-4}$ \\
B$_2$H$_4$ & $6.0\times10^{-3}$ & $2.0\times10^{-2}$ \\
C$_2$H$_4$ & $1.1\times10^{-5}$ & $7.9\times10^{-5}$ \\
C$_2$H$_6$ & $4.6\times10^{-4}$ & $3.9\times10^{-3}$ \\
PCl$_5$ & $1.6\times10^{-4}$ & $8.5\times10^{-4}$ \\
S$_8$ & $1.2\times10^{-3}$ & $7.8\times10^{-3}$ \\
CuCl$_4^{2-}$ & $2.6\times10^{-5}$ & $6.1\times10^{-5}$ \\
C$_5$H$_5^-$ & $2.7\times10^{-3}$ & $7.6\times10^{-3}$ \\
C$_6$H$_6$ & $2.5\times10^{-3}$ & $5.3\times10^{-2}$ \\
HCN & $3.3\times10^{-6}$ & $5.7\times10^{-6}$ \\
C$_2$H$_2$ & $7.8\times10^{-6}$ & $1.7\times10^{-5}$ \\
CH$_4$ & $1.9\times10^{-3}$ & $2.3\times10^{-2}$ \\
SF$_6$ & $4.2\times10^{-4}$ & $1.3\times10^{-3}$ \\
\bottomrule
\end{tabular}
\end{table}

Even though we have shown that ASyBGR preserves the underlying physics, we also tested the relative deviation between the surfaces themselves.
\Cref{tab:data2} reports the relative root-mean-square deviations (RRMSDs) between the surfaces generated with and without reduction.
The symmetry-reduced dipole moment surfaces remained in good agreement with their corresponding full surface across all test cases, with the largest RRMSD being a mere $5.2\times 10^{-2}$ a.u.
Deviations between the PESs are even smaller since the scalar value does not require an additional transformation.

\section{Conclusion}
This work presents ASyBGR, a general and robust method to reduce the computational cost of constructing a grid-based PES, or any molecular properties surface, by exploiting molecular symmetry in normal coordinates to reduce the number of grid points that must be explicitly evaluated.
ASyBGR can be broken into 10 easy steps:
\begin{enumerate}
    \item Identify as many symmetry operations as possible.
    \item Determine the symmetry group and normal coordinate irreps.
    \item Perform group closure.
    \item Produce candidate bases for each degenerate subspace.
    \item Score candidate bases and use the basis with the lowest score.
    \item Discard all coordinate-mixing operations and map the rest onto bit-strings
    \item Perform Gaussian elimination to get the bit-string basis.
    \item Identify reducible coordinates and all symmetry-valid sign change operations.
    \item Compute surface on the negative half of the reducible coordinates.
    \item Recover full surface using identified sign changes.
\end{enumerate}
Skipping or failing steps 2--6 is not fatal to ASyBGR, but may diminish its full reduction capability. 
ASyBGR does not rely on the symmetry group or irrep assignment.
Their only role is identifying degenerate subspaces during the coordinate basis optimization, which is encouraged but ultimately optional for most cases.
Even the initial group closure may fail, as closure in the bit-string representation provides a numerically forgiving fallback to make most of the detected symmetry operations.\\

Unlike current symmetry-based grid reduction methods, we are not restricted to Abelian and low-order cyclic symmetry groups,\cite{Bowman2003_533, Christiansen2019_MidasCpp} nor do we rely on brute-force sampling of the PES grid.\cite{Ziegler2018_164110}
To the best of our knowledge, a general symmetry-based grid reduction method based purely on algebraic means had not been developed yet.
Our algebraic approach maximizes the full potential that molecular symmetry has to offer while keeping computational cost low.
At a fix number of coordinates, the coordinate optimization only scales linearly with the HDMR truncation order.
Although the coordinate optimization scales exponentially with the number of degenerate subspaces, the exponential base is typically small for most chemically relevant systems.
This is an improvement over brute-force sampling where the cost scales exponentially with truncation order while the number of grid points per coordinate acts as the base.
At the limit where the number of degenerate subspaces is large, coordinate optimization may become more expensive than brute-force sampling.
In this case, coordinate optimization can be skipped entirely while retaining substantial reduction capability in most cases.\\

Although this work applied ASyBGR only to the PES and dipole moment surface within the HDMR representation, the method is applicable to any grid-based sampling of a molecular property surface. 
For example, this includes the numerical stencils used to evaluate anharmonic constants via finite difference for a Taylor series representation of the PES.\cite{Boese2005_863, Lin2008_23, Sibaev2015_2200}
More broadly, many other PES parameterization methods require sampling molecular structures to generate fitting data,\cite{Xie2010_26, Kamath2018_241702, Boussaidi2020_7598} and the same symmetry-based reduction can be used to eliminate redundant evaluations in these approaches.\\

While ASyBGR works extremely well, reducing grid sizes by as much as $81\%$, it is only valid in a normal coordinate system.
More general coordinate systems may involve linear combinations of non-degenerate normal coordinates, causing additional symmetry operations to appear coordinate-mixing in the chosen basis. 
To the best of our knowledge, a general algebraic symmetry-based grid reduction scheme for such coordinate systems has not yet been developed. 
Such a scheme may require a fundamentally different approach, particularly because some rectilinear coordinate systems exhibit different forms of symmetry, such as the permutational symmetry of localized normal coordinates.\cite{Ziegler2018_164110, Ziegler2019_4187} 
Extending symmetry-based grid reduction beyond normal coordinates therefore remains an important direction for future work.

\begin{acknowledgement}
	This work was financially supported by the Swiss National Science Foundation (Grant No. SNF 200021\_219616).
\end{acknowledgement}






\begin{mcitethebibliography}{75}
\providecommand*\natexlab[1]{#1}
\providecommand*\mciteSetBstSublistMode[1]{}
\providecommand*\mciteSetBstMaxWidthForm[2]{}
\providecommand*\mciteBstWouldAddEndPuncttrue
  {\def\EndOfBibitem{\unskip.}}
\providecommand*\mciteBstWouldAddEndPunctfalse
  {\let\EndOfBibitem\relax}
\providecommand*\mciteSetBstMidEndSepPunct[3]{}
\providecommand*\mciteSetBstSublistLabelBeginEnd[3]{}
\providecommand*\EndOfBibitem{}
\mciteSetBstSublistMode{f}
\mciteSetBstMaxWidthForm{subitem}{(\alph{mcitesubitemcount})}
\mciteSetBstSublistLabelBeginEnd
  {\mcitemaxwidthsubitemform\space}
  {\relax}
  {\relax}

\bibitem[Bowman(1986)]{Bowman1986_202}
Bowman,~J.~M. The self-consistent-field approach to polyatomic vibrations.
  \emph{Acc. Chem. Res.} \textbf{1986}, \emph{19}, 202--208\relax
\mciteBstWouldAddEndPuncttrue
\mciteSetBstMidEndSepPunct{\mcitedefaultmidpunct}
{\mcitedefaultendpunct}{\mcitedefaultseppunct}\relax
\EndOfBibitem
\bibitem[Meyer \latin{et~al.}(1990)Meyer, Manthe, and Cederbaum]{Meyer1990_73}
Meyer,~H.-D.; Manthe,~U.; Cederbaum,~L.~S. The multi-configurational
  time-dependent {Hartree} approach. \emph{Chem. Phys. Lett.} \textbf{1990},
  \emph{165}, 73--78\relax
\mciteBstWouldAddEndPuncttrue
\mciteSetBstMidEndSepPunct{\mcitedefaultmidpunct}
{\mcitedefaultendpunct}{\mcitedefaultseppunct}\relax
\EndOfBibitem
\bibitem[Carter \latin{et~al.}(1997)Carter, Culik, and
  Bowman]{Carter1997_10458}
Carter,~S.; Culik,~S.~J.; Bowman,~J.~M. Vibrational self-consistent field
  method for many-mode systems: A new approach and application to the
  vibrations of {CO} adsorbed on {Cu}(100). \emph{J. Chem. Phys.}
  \textbf{1997}, \emph{107}, 10458--10469\relax
\mciteBstWouldAddEndPuncttrue
\mciteSetBstMidEndSepPunct{\mcitedefaultmidpunct}
{\mcitedefaultendpunct}{\mcitedefaultseppunct}\relax
\EndOfBibitem
\bibitem[Beck \latin{et~al.}(2000)Beck, J{\"a}ckle, Worth, and
  Meyer]{Beck2000_1}
Beck,~M.~H.; J{\"a}ckle,~A.; Worth,~G.~A.; Meyer,~H.-D. The multiconfiguration
  time-dependent {Hartree} ({MCTDH}) method: A highly efficient algorithm for
  propagating wavepackets. \emph{Phys. Rep.} \textbf{2000}, \emph{324},
  1--105\relax
\mciteBstWouldAddEndPuncttrue
\mciteSetBstMidEndSepPunct{\mcitedefaultmidpunct}
{\mcitedefaultendpunct}{\mcitedefaultseppunct}\relax
\EndOfBibitem
\bibitem[Bowman \latin{et~al.}(2003)Bowman, Carter, and Huang]{Bowman2003_533}
Bowman,~J.~M.; Carter,~S.; Huang,~X. {MULTIMODE}: A code to calculate
  rovibrational energies of polyatomic molecules. \emph{Int. Rev. Phys. Chem.}
  \textbf{2003}, \emph{22}, 533--549\relax
\mciteBstWouldAddEndPuncttrue
\mciteSetBstMidEndSepPunct{\mcitedefaultmidpunct}
{\mcitedefaultendpunct}{\mcitedefaultseppunct}\relax
\EndOfBibitem
\bibitem[Christiansen(2003)]{Christiansen2003_5773}
Christiansen,~O. M{\o}ller--Plesset perturbation theory for vibrational wave
  functions. \emph{J. Chem. Phys.} \textbf{2003}, \emph{119}, 5773--5781\relax
\mciteBstWouldAddEndPuncttrue
\mciteSetBstMidEndSepPunct{\mcitedefaultmidpunct}
{\mcitedefaultendpunct}{\mcitedefaultseppunct}\relax
\EndOfBibitem
\bibitem[Christiansen(2004)]{Christiansen2004_2140}
Christiansen,~O. A second quantization formulation of multimode dynamics.
  \emph{J. Chem. Phys.} \textbf{2004}, \emph{120}, 2140--2148\relax
\mciteBstWouldAddEndPuncttrue
\mciteSetBstMidEndSepPunct{\mcitedefaultmidpunct}
{\mcitedefaultendpunct}{\mcitedefaultseppunct}\relax
\EndOfBibitem
\bibitem[Christiansen(2004)]{Christiansen2004_2149}
Christiansen,~O. Vibrational coupled cluster theory. \emph{J. Chem. Phys.}
  \textbf{2004}, \emph{120}, 2149--2159\relax
\mciteBstWouldAddEndPuncttrue
\mciteSetBstMidEndSepPunct{\mcitedefaultmidpunct}
{\mcitedefaultendpunct}{\mcitedefaultseppunct}\relax
\EndOfBibitem
\bibitem[Boese \latin{et~al.}(2005)Boese, Klopper, and Martin]{Boese2005_863}
Boese,~A.~D.; Klopper,~W.; Martin,~J. M.~L. Anharmonic force fields and
  thermodynamic functions using density functional theory. \emph{Mol. Phys.}
  \textbf{2005}, \emph{103}, 863--876\relax
\mciteBstWouldAddEndPuncttrue
\mciteSetBstMidEndSepPunct{\mcitedefaultmidpunct}
{\mcitedefaultendpunct}{\mcitedefaultseppunct}\relax
\EndOfBibitem
\bibitem[Christiansen(2005)]{Christiansen2005_194105}
Christiansen,~O. Response theory for vibrational wave functions. \emph{J. Chem.
  Phys.} \textbf{2005}, \emph{122}, 194105\relax
\mciteBstWouldAddEndPuncttrue
\mciteSetBstMidEndSepPunct{\mcitedefaultmidpunct}
{\mcitedefaultendpunct}{\mcitedefaultseppunct}\relax
\EndOfBibitem
\bibitem[Rauhut(2007)]{Rauhut2007_184109}
Rauhut,~G. Configuration selection as a route towards efficient vibrational
  configuration interaction calculations. \emph{J. Chem. Phys.} \textbf{2007},
  \emph{127}, 184109\relax
\mciteBstWouldAddEndPuncttrue
\mciteSetBstMidEndSepPunct{\mcitedefaultmidpunct}
{\mcitedefaultendpunct}{\mcitedefaultseppunct}\relax
\EndOfBibitem
\bibitem[Bowman \latin{et~al.}(2008)Bowman, Carrington, and
  Meyer]{Bowman2008_2145}
Bowman,~J.~M.; Carrington,~T.; Meyer,~H.-D. Variational quantum approaches for
  computing vibrational energies of polyatomic molecules. \emph{Mol. Phys.}
  \textbf{2008}, \emph{106}, 2145--2182\relax
\mciteBstWouldAddEndPuncttrue
\mciteSetBstMidEndSepPunct{\mcitedefaultmidpunct}
{\mcitedefaultendpunct}{\mcitedefaultseppunct}\relax
\EndOfBibitem
\bibitem[Petit and McCoy(2013)Petit, and McCoy]{Petit2013_7009}
Petit,~A.~S.; McCoy,~A.~B. Diffusion {Monte Carlo} in Internal Coordinates.
  \emph{J. Phys. Chem. A} \textbf{2013}, \emph{117}, 7009--7018\relax
\mciteBstWouldAddEndPuncttrue
\mciteSetBstMidEndSepPunct{\mcitedefaultmidpunct}
{\mcitedefaultendpunct}{\mcitedefaultseppunct}\relax
\EndOfBibitem
\bibitem[Garnier \latin{et~al.}(2016)Garnier, Odunlami, Le~Bris, B{\'e}gu{\'e},
  Baraille, and Coulaud]{Garnier2016_204123}
Garnier,~R.; Odunlami,~M.; Le~Bris,~V.; B{\'e}gu{\'e},~D.; Baraille,~I.;
  Coulaud,~O. Adaptive vibrational configuration interaction ({A-VCI}): A
  posteriori error estimation to efficiently compute anharmonic {IR} spectra.
  \emph{J. Chem. Phys.} \textbf{2016}, \emph{144}, 204123\relax
\mciteBstWouldAddEndPuncttrue
\mciteSetBstMidEndSepPunct{\mcitedefaultmidpunct}
{\mcitedefaultendpunct}{\mcitedefaultseppunct}\relax
\EndOfBibitem
\bibitem[Baiardi \latin{et~al.}(2017)Baiardi, Stein, Barone, and
  Reiher]{Baiardi2017_3764}
Baiardi,~A.; Stein,~C.~J.; Barone,~V.; Reiher,~M. Vibrational Density Matrix
  Renormalization Group. \emph{J. Chem. Theory Comput.} \textbf{2017},
  \emph{13}, 3764--3777\relax
\mciteBstWouldAddEndPuncttrue
\mciteSetBstMidEndSepPunct{\mcitedefaultmidpunct}
{\mcitedefaultendpunct}{\mcitedefaultseppunct}\relax
\EndOfBibitem
\bibitem[Christiansen \latin{et~al.}(2019)Christiansen, Artiukhin, Godtliebsen,
  Gras, Gy{\H{o}}rffy, Hansen, Hansen, Klinting, Kongsted, K{\"o}nig, Madsen,
  Madsen, Monrad, Schmitz, Seidler, Sneskov, Sparta, Thomsen, Toffoli, and
  Zoccante]{Christiansen2019_MidasCpp}
Christiansen,~O.; Artiukhin,~D.; Godtliebsen,~I.~H.; Gras,~E.~M.;
  Gy{\H{o}}rffy,~W.; Hansen,~M.~B.; Hansen,~M.~B.; Klinting,~E.~L.;
  Kongsted,~J.; K{\"o}nig,~C.; Madsen,~D.; Madsen,~N.~K.; Monrad,~K.;
  Schmitz,~G.; Seidler,~P.; Sneskov,~K.; Sparta,~M.; Thomsen,~B.; Toffoli,~D.;
  Zoccante,~A. MidasCpp: Molecular Interactions Dynamics And Simulation
  Chemistry Program Package. 2019;
  \url{https://source.coderefinery.org/midascpp/midascpp}\relax
\mciteBstWouldAddEndPuncttrue
\mciteSetBstMidEndSepPunct{\mcitedefaultmidpunct}
{\mcitedefaultendpunct}{\mcitedefaultseppunct}\relax
\EndOfBibitem
\bibitem[Larsson(2019)]{Larsson2019_204102}
Larsson,~H.~R. Computing vibrational eigenstates with tree tensor network
  states ({TTNS}). \emph{J. Chem. Phys.} \textbf{2019}, \emph{151},
  204102\relax
\mciteBstWouldAddEndPuncttrue
\mciteSetBstMidEndSepPunct{\mcitedefaultmidpunct}
{\mcitedefaultendpunct}{\mcitedefaultseppunct}\relax
\EndOfBibitem
\bibitem[Fetherolf and Berkelbach(2021)Fetherolf, and
  Berkelbach]{Fetherolf2021_074104}
Fetherolf,~J.~H.; Berkelbach,~T.~C. Vibrational heat-bath configuration
  interaction. \emph{J. Chem. Phys.} \textbf{2021}, \emph{154}, 074104\relax
\mciteBstWouldAddEndPuncttrue
\mciteSetBstMidEndSepPunct{\mcitedefaultmidpunct}
{\mcitedefaultendpunct}{\mcitedefaultseppunct}\relax
\EndOfBibitem
\bibitem[Bowman(2022)]{Bowman2022_Book}
Bowman,~J.~M., Ed. \emph{Vibrational Dynamics of Molecules}; World Scientific,
  2022\relax
\mciteBstWouldAddEndPuncttrue
\mciteSetBstMidEndSepPunct{\mcitedefaultmidpunct}
{\mcitedefaultendpunct}{\mcitedefaultseppunct}\relax
\EndOfBibitem
\bibitem[Glaser \latin{et~al.}(2023)Glaser, Baiardi, and
  Reiher]{Glaser2023_9329}
Glaser,~N.; Baiardi,~A.; Reiher,~M. Flexible {DMRG}-Based Framework for
  Anharmonic Vibrational Calculations. \emph{J. Chem. Theory Comput.}
  \textbf{2023}, \emph{19}, 9329--9343\relax
\mciteBstWouldAddEndPuncttrue
\mciteSetBstMidEndSepPunct{\mcitedefaultmidpunct}
{\mcitedefaultendpunct}{\mcitedefaultseppunct}\relax
\EndOfBibitem
\bibitem[Henry and Amat(1965)Henry, and Amat]{Henry1965_168}
Henry,~L.; Amat,~G. The quartic anharmonic potential function of polyatomic
  molecules. \emph{J. Mol. Spectrosc.} \textbf{1965}, \emph{15}, 168--179\relax
\mciteBstWouldAddEndPuncttrue
\mciteSetBstMidEndSepPunct{\mcitedefaultmidpunct}
{\mcitedefaultendpunct}{\mcitedefaultseppunct}\relax
\EndOfBibitem
\bibitem[Henry and Amat(1961)Henry, and Amat]{Henry1961_319}
Henry,~L.; Amat,~G. The cubic anharmonic potential function of polyatomic
  molecules. \emph{J. Mol. Spectrosc.} \textbf{1961}, \emph{5}, 319--325\relax
\mciteBstWouldAddEndPuncttrue
\mciteSetBstMidEndSepPunct{\mcitedefaultmidpunct}
{\mcitedefaultendpunct}{\mcitedefaultseppunct}\relax
\EndOfBibitem
\bibitem[J{\"a}ckle and Meyer(1996)J{\"a}ckle, and Meyer]{Jackle1996_7974}
J{\"a}ckle,~A.; Meyer,~H.-D. Product representation of potential energy
  surfaces. \emph{J. Chem. Phys.} \textbf{1996}, \emph{104}, 7974--7984\relax
\mciteBstWouldAddEndPuncttrue
\mciteSetBstMidEndSepPunct{\mcitedefaultmidpunct}
{\mcitedefaultendpunct}{\mcitedefaultseppunct}\relax
\EndOfBibitem
\bibitem[Li \latin{et~al.}(2001)Li, Rosenthal, and Rabitz]{Li2001_7765}
Li,~G.; Rosenthal,~C.; Rabitz,~H. High Dimensional Model Representations.
  \emph{J. Phys. Chem. A} \textbf{2001}, \emph{105}, 7765--7777\relax
\mciteBstWouldAddEndPuncttrue
\mciteSetBstMidEndSepPunct{\mcitedefaultmidpunct}
{\mcitedefaultendpunct}{\mcitedefaultseppunct}\relax
\EndOfBibitem
\bibitem[Rauhut(2004)]{Rauhut2004_9313}
Rauhut,~G. Efficient calculation of potential energy surfaces for the
  generation of vibrational wave functions. \emph{J. Chem. Phys.}
  \textbf{2004}, \emph{121}, 9313--9322\relax
\mciteBstWouldAddEndPuncttrue
\mciteSetBstMidEndSepPunct{\mcitedefaultmidpunct}
{\mcitedefaultendpunct}{\mcitedefaultseppunct}\relax
\EndOfBibitem
\bibitem[Manzhos and Carrington(2006)Manzhos, and
  Carrington]{Manzhos2006_084109}
Manzhos,~S.; Carrington,~T.,~Jr. A random-sampling high dimensional model
  representation neural network for building potential energy surfaces.
  \emph{J. Chem. Phys.} \textbf{2006}, \emph{125}, 084109\relax
\mciteBstWouldAddEndPuncttrue
\mciteSetBstMidEndSepPunct{\mcitedefaultmidpunct}
{\mcitedefaultendpunct}{\mcitedefaultseppunct}\relax
\EndOfBibitem
\bibitem[Manzhos and Carrington(2008)Manzhos, and
  Carrington]{Manzhos2008_224104}
Manzhos,~S.; Carrington,~T.,~Jr. Using neural networks, optimized coordinates,
  and high-dimensional model representations to obtain a vinyl bromide
  potential surface. \emph{J. Chem. Phys.} \textbf{2008}, \emph{129},
  224104\relax
\mciteBstWouldAddEndPuncttrue
\mciteSetBstMidEndSepPunct{\mcitedefaultmidpunct}
{\mcitedefaultendpunct}{\mcitedefaultseppunct}\relax
\EndOfBibitem
\bibitem[Lin \latin{et~al.}(2008)Lin, Gilbert, and Gill]{Lin2008_23}
Lin,~C.~Y.; Gilbert,~A. T.~B.; Gill,~P. M.~W. Calculating molecular vibrational
  spectra beyond the harmonic approximation. \emph{Theor. Chem. Acc.}
  \textbf{2008}, \emph{120}, 23--35\relax
\mciteBstWouldAddEndPuncttrue
\mciteSetBstMidEndSepPunct{\mcitedefaultmidpunct}
{\mcitedefaultendpunct}{\mcitedefaultseppunct}\relax
\EndOfBibitem
\bibitem[Sparta \latin{et~al.}(2009)Sparta, H{\o}yvik, Toffoli, and
  Christiansen]{Sparta2009_8712}
Sparta,~M.; H{\o}yvik,~I.-M.; Toffoli,~D.; Christiansen,~O. Potential Energy
  Surfaces for Vibrational Structure Calculations from a Multiresolution
  Adaptive Density-Guided Approach: Implementation and Test Calculations.
  \emph{J. Phys. Chem. A} \textbf{2009}, \emph{113}, 8712--8723\relax
\mciteBstWouldAddEndPuncttrue
\mciteSetBstMidEndSepPunct{\mcitedefaultmidpunct}
{\mcitedefaultendpunct}{\mcitedefaultseppunct}\relax
\EndOfBibitem
\bibitem[Sparta \latin{et~al.}(2010)Sparta, Hansen, Matito, Toffoli, and
  Christiansen]{Sparta2010_3162}
Sparta,~M.; Hansen,~M.~B.; Matito,~E.; Toffoli,~D.; Christiansen,~O. Using
  Electronic Energy Derivative Information in Automated Potential Energy
  Surface Construction for Vibrational Calculations. \emph{J. Chem. Theory
  Comput.} \textbf{2010}, \emph{6}, 3162--3175\relax
\mciteBstWouldAddEndPuncttrue
\mciteSetBstMidEndSepPunct{\mcitedefaultmidpunct}
{\mcitedefaultendpunct}{\mcitedefaultseppunct}\relax
\EndOfBibitem
\bibitem[Xie and Bowman(2010)Xie, and Bowman]{Xie2010_26}
Xie,~Z.; Bowman,~J.~M. Permutationally Invariant Polynomial Basis for Molecular
  Energy Surface Fitting via Monomial Symmetrization. \emph{J. Chem. Theory
  Comput.} \textbf{2010}, \emph{6}, 26--34\relax
\mciteBstWouldAddEndPuncttrue
\mciteSetBstMidEndSepPunct{\mcitedefaultmidpunct}
{\mcitedefaultendpunct}{\mcitedefaultseppunct}\relax
\EndOfBibitem
\bibitem[Cs{\'a}sz{\'a}r(2012)]{Csaszar2012_273}
Cs{\'a}sz{\'a}r,~A.~G. Anharmonic molecular force fields. \emph{Wiley
  Interdiscip. Rev. Comput. Mol. Sci.} \textbf{2012}, \emph{2}, 273--289\relax
\mciteBstWouldAddEndPuncttrue
\mciteSetBstMidEndSepPunct{\mcitedefaultmidpunct}
{\mcitedefaultendpunct}{\mcitedefaultseppunct}\relax
\EndOfBibitem
\bibitem[Sibaev and Crittenden(2015)Sibaev, and Crittenden]{Sibaev2015_2200}
Sibaev,~M.; Crittenden,~D.~L. The {PyPES} library of high quality semi-global
  potential energy surfaces. \emph{J. Comput. Chem.} \textbf{2015}, \emph{36},
  2200--2207\relax
\mciteBstWouldAddEndPuncttrue
\mciteSetBstMidEndSepPunct{\mcitedefaultmidpunct}
{\mcitedefaultendpunct}{\mcitedefaultseppunct}\relax
\EndOfBibitem
\bibitem[Avila and Carrington(2015)Avila, and Carrington]{Avila2015_044106}
Avila,~G.; Carrington,~T.,~Jr. Using multi-dimensional Smolyak interpolation to
  make a sum-of-products potential. \emph{J. Chem. Phys.} \textbf{2015},
  \emph{143}, 044106\relax
\mciteBstWouldAddEndPuncttrue
\mciteSetBstMidEndSepPunct{\mcitedefaultmidpunct}
{\mcitedefaultendpunct}{\mcitedefaultseppunct}\relax
\EndOfBibitem
\bibitem[Ziegler and Rauhut(2016)Ziegler, and Rauhut]{Ziegler2016_114114}
Ziegler,~B.; Rauhut,~G. Efficient generation of sum-of-products representations
  of high-dimensional potential energy surfaces based on multimode expansions.
  \emph{J. Chem. Phys.} \textbf{2016}, \emph{144}, 114114\relax
\mciteBstWouldAddEndPuncttrue
\mciteSetBstMidEndSepPunct{\mcitedefaultmidpunct}
{\mcitedefaultendpunct}{\mcitedefaultseppunct}\relax
\EndOfBibitem
\bibitem[Tan and Kuo(2018)Tan, and Kuo]{Tan2018_6405}
Tan,~J.~A.; Kuo,~J.-L. Multilevel Approach for Direct {VSCF/VCI MULTIMODE}
  Calculations with Applications to Large ``Zundel'' Cations. \emph{J. Chem.
  Theory Comput.} \textbf{2018}, \emph{14}, 6405--6416\relax
\mciteBstWouldAddEndPuncttrue
\mciteSetBstMidEndSepPunct{\mcitedefaultmidpunct}
{\mcitedefaultendpunct}{\mcitedefaultseppunct}\relax
\EndOfBibitem
\bibitem[Boussaidi \latin{et~al.}(2020)Boussaidi, Ren, Voytsekhovsky, and
  Manzhos]{Boussaidi2020_7598}
Boussaidi,~M.~A.; Ren,~O.; Voytsekhovsky,~D.; Manzhos,~S. Random Sampling High
  Dimensional Model Representation Gaussian Process Regression ({RS-HDMR-GPR})
  for Multivariate Function Representation: Application to Molecular Potential
  Energy Surfaces. \emph{J. Phys. Chem. A} \textbf{2020}, \emph{124},
  7598--7607\relax
\mciteBstWouldAddEndPuncttrue
\mciteSetBstMidEndSepPunct{\mcitedefaultmidpunct}
{\mcitedefaultendpunct}{\mcitedefaultseppunct}\relax
\EndOfBibitem
\bibitem[Yagi \latin{et~al.}(2000)Yagi, Taketsugu, Hirao, and
  Gordon]{Yagi2000_1005}
Yagi,~K.; Taketsugu,~T.; Hirao,~K.; Gordon,~M.~S. Direct vibrational
  self-consistent field method: Applications to H$_2$O and H$_2$CO. \emph{J.
  Chem. Phys.} \textbf{2000}, \emph{113}, 1005--1017\relax
\mciteBstWouldAddEndPuncttrue
\mciteSetBstMidEndSepPunct{\mcitedefaultmidpunct}
{\mcitedefaultendpunct}{\mcitedefaultseppunct}\relax
\EndOfBibitem
\bibitem[Huang \latin{et~al.}(2002)Huang, Carter, and Bowman]{Huang2002_8182}
Huang,~X.; Carter,~S.; Bowman,~J.~M. Ab Initio Potential Energy Surface and
  Vibrational Energies of H$_3$O$^+$ and Its Isotopomers. \emph{J. Phys. Chem.
  B} \textbf{2002}, \emph{106}, 8182--8188\relax
\mciteBstWouldAddEndPuncttrue
\mciteSetBstMidEndSepPunct{\mcitedefaultmidpunct}
{\mcitedefaultendpunct}{\mcitedefaultseppunct}\relax
\EndOfBibitem
\bibitem[Wang and Carrington(2003)Wang, and Carrington]{Wang2003_94}
Wang,~X.-G.; Carrington,~J.,~Tucker Using $C_{3v}$ symmetry with polyspherical
  coordinates for methane. \emph{J. Chem. Phys.} \textbf{2003}, \emph{119},
  94--100\relax
\mciteBstWouldAddEndPuncttrue
\mciteSetBstMidEndSepPunct{\mcitedefaultmidpunct}
{\mcitedefaultendpunct}{\mcitedefaultseppunct}\relax
\EndOfBibitem
\bibitem[Wang and Carrington(2005)Wang, and Carrington]{Wang2005_154303}
Wang,~X.-G.; Carrington,~T.,~Jr. Improving the calculation of rovibrational
  spectra of five-atom molecules with three identical atoms by using a
  $C_{3v}(G_6)$ symmetry-adapted grid: Applied to CH$_3$D and CHD$_3$. \emph{J.
  Chem. Phys.} \textbf{2005}, \emph{123}, 154303\relax
\mciteBstWouldAddEndPuncttrue
\mciteSetBstMidEndSepPunct{\mcitedefaultmidpunct}
{\mcitedefaultendpunct}{\mcitedefaultseppunct}\relax
\EndOfBibitem
\bibitem[Feller and Peterson(2009)Feller, and Peterson]{Feller2009_154306}
Feller,~D.; Peterson,~K.~A. High level coupled cluster determination of the
  structure, frequencies, and heat of formation of water. \emph{J. Chem. Phys.}
  \textbf{2009}, \emph{131}, 154306\relax
\mciteBstWouldAddEndPuncttrue
\mciteSetBstMidEndSepPunct{\mcitedefaultmidpunct}
{\mcitedefaultendpunct}{\mcitedefaultseppunct}\relax
\EndOfBibitem
\bibitem[Pradhan \latin{et~al.}(2013)Pradhan, Carre{\'o}n-Macedo, Cuervo,
  Schr{\"o}der, and Brown]{Pradhan2013_6925}
Pradhan,~E.; Carre{\'o}n-Macedo,~J.-L.; Cuervo,~J.~E.; Schr{\"o}der,~M.;
  Brown,~A. Ab Initio Potential Energy and Dipole Moment Surfaces for CS$_2$:
  Determination of Molecular Vibrational Energies. \emph{J. Phys. Chem. A}
  \textbf{2013}, \emph{117}, 6925--6931\relax
\mciteBstWouldAddEndPuncttrue
\mciteSetBstMidEndSepPunct{\mcitedefaultmidpunct}
{\mcitedefaultendpunct}{\mcitedefaultseppunct}\relax
\EndOfBibitem
\bibitem[Oschetzki and Rauhut(2014)Oschetzki, and Rauhut]{Oschetzki2014_16426}
Oschetzki,~D.; Rauhut,~G. Pushing the limits in accurate vibrational structure
  calculations: anharmonic frequencies of lithium fluoride clusters (LiF)$_n$,
  $n$ = 2--10. \emph{Phys. Chem. Chem. Phys.} \textbf{2014}, \emph{16},
  16426--16435\relax
\mciteBstWouldAddEndPuncttrue
\mciteSetBstMidEndSepPunct{\mcitedefaultmidpunct}
{\mcitedefaultendpunct}{\mcitedefaultseppunct}\relax
\EndOfBibitem
\bibitem[Nikitin \latin{et~al.}(2016)Nikitin, Rey, and
  Tyuterev]{Nikitin2016_114309}
Nikitin,~A.~V.; Rey,~M.; Tyuterev,~V.~G. First fully ab initio potential energy
  surface of methane with a spectroscopic accuracy. \emph{J. Chem. Phys.}
  \textbf{2016}, \emph{145}, 114309\relax
\mciteBstWouldAddEndPuncttrue
\mciteSetBstMidEndSepPunct{\mcitedefaultmidpunct}
{\mcitedefaultendpunct}{\mcitedefaultseppunct}\relax
\EndOfBibitem
\bibitem[Ziegler and Rauhut(2018)Ziegler, and Rauhut]{Ziegler2018_164110}
Ziegler,~B.; Rauhut,~G. Rigorous use of symmetry within the construction of
  multidimensional potential energy surfaces. \emph{J. Chem. Phys.}
  \textbf{2018}, \emph{149}, 164110\relax
\mciteBstWouldAddEndPuncttrue
\mciteSetBstMidEndSepPunct{\mcitedefaultmidpunct}
{\mcitedefaultendpunct}{\mcitedefaultseppunct}\relax
\EndOfBibitem
\bibitem[Mitoli \latin{et~al.}(2023)Mitoli, Maul, and Erba]{Mitoli2023_3671}
Mitoli,~D.; Maul,~J.; Erba,~A. Anharmonic Terms of the Potential Energy
  Surface: A Group Theoretical Approach. \emph{Cryst. Growth Des.}
  \textbf{2023}, \emph{23}, 3671--3680\relax
\mciteBstWouldAddEndPuncttrue
\mciteSetBstMidEndSepPunct{\mcitedefaultmidpunct}
{\mcitedefaultendpunct}{\mcitedefaultseppunct}\relax
\EndOfBibitem
\bibitem[Seko and Togo(2024)Seko, and Togo]{Seko2024_214302}
Seko,~A.; Togo,~A. Projector-based efficient estimation of force constants.
  \emph{Phys. Rev. B} \textbf{2024}, \emph{110}, 214302\relax
\mciteBstWouldAddEndPuncttrue
\mciteSetBstMidEndSepPunct{\mcitedefaultmidpunct}
{\mcitedefaultendpunct}{\mcitedefaultseppunct}\relax
\EndOfBibitem
\bibitem[Schneider and Rauhut(2024)Schneider, and Rauhut]{Schneider2024_094102}
Schneider,~M.; Rauhut,~G. Comparison of curvilinear coordinates within
  vibrational structure calculations based on automatically generated potential
  energy surfaces. \emph{J. Chem. Phys.} \textbf{2024}, \emph{161},
  094102\relax
\mciteBstWouldAddEndPuncttrue
\mciteSetBstMidEndSepPunct{\mcitedefaultmidpunct}
{\mcitedefaultendpunct}{\mcitedefaultseppunct}\relax
\EndOfBibitem
\bibitem[Altmann(1986)]{Altmann1986_Book}
Altmann,~S.~L. \emph{Rotations, Quaternions, and Double Groups}; Clarendon
  Press: Oxford, 1986\relax
\mciteBstWouldAddEndPuncttrue
\mciteSetBstMidEndSepPunct{\mcitedefaultmidpunct}
{\mcitedefaultendpunct}{\mcitedefaultseppunct}\relax
\EndOfBibitem
\bibitem[Altmann and Herzig(1994)Altmann, and Herzig]{Altmann1994_Book}
Altmann,~S.~L.; Herzig,~P. \emph{Point-Group Theory Tables}; Clarendon Press,
  1994\relax
\mciteBstWouldAddEndPuncttrue
\mciteSetBstMidEndSepPunct{\mcitedefaultmidpunct}
{\mcitedefaultendpunct}{\mcitedefaultseppunct}\relax
\EndOfBibitem
\bibitem[Knowles(2022)]{Knowles2022_161}
Knowles,~P.~J. The determination of point groups from imprecise molecular
  geometries. \emph{J. Math. Chem.} \textbf{2022}, \emph{60}, 161--171\relax
\mciteBstWouldAddEndPuncttrue
\mciteSetBstMidEndSepPunct{\mcitedefaultmidpunct}
{\mcitedefaultendpunct}{\mcitedefaultseppunct}\relax
\EndOfBibitem
\bibitem[Gunde \latin{et~al.}(2024)Gunde, Salles, Grisanti, Martin-Samos, and
  Hemeryck]{Gunde2024_062503}
Gunde,~M.; Salles,~N.; Grisanti,~L.; Martin-Samos,~L.; Hemeryck,~A. {SOFI}:
  Finding point group symmetries in atomic clusters as finding the set of
  degenerate solutions in a shape-matching problem. \emph{J. Chem. Phys.}
  \textbf{2024}, \emph{161}, 062503\relax
\mciteBstWouldAddEndPuncttrue
\mciteSetBstMidEndSepPunct{\mcitedefaultmidpunct}
{\mcitedefaultendpunct}{\mcitedefaultseppunct}\relax
\EndOfBibitem
\bibitem[Budai \latin{et~al.}(1977)Budai, Kovrikov, Lyudchik, Popov, and
  Umreiko]{Budai1977_97}
Budai,~L.~I.; Kovrikov,~A.~B.; Lyudchik,~A.~M.; Popov,~V.~G.; Umreiko,~D.~S.
  Automatic symmetry analysis in the vibrational and electronic spectroscopy of
  molecules. \emph{J. Appl. Spectrosc.} \textbf{1977}, \emph{26}, 97--100\relax
\mciteBstWouldAddEndPuncttrue
\mciteSetBstMidEndSepPunct{\mcitedefaultmidpunct}
{\mcitedefaultendpunct}{\mcitedefaultseppunct}\relax
\EndOfBibitem
\bibitem[Zabrodsky \latin{et~al.}(1992)Zabrodsky, Peleg, and
  Avnir]{Zabrodsky1992_7843}
Zabrodsky,~H.; Peleg,~S.; Avnir,~D. Continuous symmetry measures. \emph{J. Am.
  Chem. Soc.} \textbf{1992}, \emph{114}, 7843--7851\relax
\mciteBstWouldAddEndPuncttrue
\mciteSetBstMidEndSepPunct{\mcitedefaultmidpunct}
{\mcitedefaultendpunct}{\mcitedefaultseppunct}\relax
\EndOfBibitem
\bibitem[Pilati and Forni(1998)Pilati, and Forni]{Pilati1998_503}
Pilati,~T.; Forni,~A. SYMMOL: a program to find the maximum symmetry group in
  an atom cluster, given a prefixed tolerance. \emph{J. Appl. Cryst.}
  \textbf{1998}, \emph{31}, 503--504\relax
\mciteBstWouldAddEndPuncttrue
\mciteSetBstMidEndSepPunct{\mcitedefaultmidpunct}
{\mcitedefaultendpunct}{\mcitedefaultseppunct}\relax
\EndOfBibitem
\bibitem[Ivanov and Sch{\"u}{\"u}rmann(1999)Ivanov, and
  Sch{\"u}{\"u}rmann]{Ivanov1999_728}
Ivanov,~J.; Sch{\"u}{\"u}rmann,~G. Simple Algorithms for Determining the
  Molecular Symmetry. \emph{J. Chem. Inf. Comput. Sci.} \textbf{1999},
  \emph{39}, 728--737\relax
\mciteBstWouldAddEndPuncttrue
\mciteSetBstMidEndSepPunct{\mcitedefaultmidpunct}
{\mcitedefaultendpunct}{\mcitedefaultseppunct}\relax
\EndOfBibitem
\bibitem[Largent \latin{et~al.}(2012)Largent, Polik, and
  Schmidt]{Largent2012_1637}
Largent,~R.~J.; Polik,~W.~F.; Schmidt,~J.~R. Symmetrizer: Algorithmic
  determination of point groups in nearly symmetric molecules. \emph{J. Comput.
  Chem.} \textbf{2012}, \emph{33}, 1637--1642\relax
\mciteBstWouldAddEndPuncttrue
\mciteSetBstMidEndSepPunct{\mcitedefaultmidpunct}
{\mcitedefaultendpunct}{\mcitedefaultseppunct}\relax
\EndOfBibitem
\bibitem[Johansson and Veryazov(2017)Johansson, and Veryazov]{Johansson2017_8}
Johansson,~M.; Veryazov,~V. Automatic procedure for generating symmetry adapted
  wavefunctions. \emph{J. Cheminform.} \textbf{2017}, \emph{9}, 8\relax
\mciteBstWouldAddEndPuncttrue
\mciteSetBstMidEndSepPunct{\mcitedefaultmidpunct}
{\mcitedefaultendpunct}{\mcitedefaultseppunct}\relax
\EndOfBibitem
\bibitem[Gyevi-Nagy and Tasi(2017)Gyevi-Nagy, and Tasi]{GyeviNagy2017_156}
Gyevi-Nagy,~L.; Tasi,~G. SYVA: A program to analyze symmetry of molecules based
  on vector algebra. \emph{Comput. Phys. Commun.} \textbf{2017}, \emph{215},
  156--164\relax
\mciteBstWouldAddEndPuncttrue
\mciteSetBstMidEndSepPunct{\mcitedefaultmidpunct}
{\mcitedefaultendpunct}{\mcitedefaultseppunct}\relax
\EndOfBibitem
\bibitem[Huynh \latin{et~al.}(2024)Huynh, Wibowo-Teale, and
  Wibowo-Teale]{Huynh2024_114}
Huynh,~B.~C.; Wibowo-Teale,~M.; Wibowo-Teale,~A.~M. {QSym}$^2$: A Quantum
  Symbolic Symmetry Analysis Program for Electronic Structure. \emph{J. Chem.
  Theory Comput.} \textbf{2024}, \emph{20}, 114--133\relax
\mciteBstWouldAddEndPuncttrue
\mciteSetBstMidEndSepPunct{\mcitedefaultmidpunct}
{\mcitedefaultendpunct}{\mcitedefaultseppunct}\relax
\EndOfBibitem
\bibitem[Nielsen \latin{et~al.}(2024)Nielsen, Le~Guennic, and
  S{\o}rensen]{Nielsen2024_5740}
Nielsen,~V. R.~M.; Le~Guennic,~B.; S{\o}rensen,~T.~J. Evaluation of Point Group
  Symmetry in Lanthanide(III) Complexes: A New Implementation of a Continuous
  Symmetry Operation Measure with Autonomous Assignment of the Principal Axis.
  \emph{J. Phys. Chem. A} \textbf{2024}, \emph{128}, 5740--5751\relax
\mciteBstWouldAddEndPuncttrue
\mciteSetBstMidEndSepPunct{\mcitedefaultmidpunct}
{\mcitedefaultendpunct}{\mcitedefaultseppunct}\relax
\EndOfBibitem
\bibitem[Nielsen and S{\o}rensen(2025)Nielsen, and
  S{\o}rensen]{Nielsen2025_11122}
Nielsen,~V. R.~M.; S{\o}rensen,~T.~J. Determining molecular structure,
  coordination geometry, and molecular symmetry using a continuous symmetry
  operation measure software. \emph{Nat. Commun.} \textbf{2025}, \emph{16},
  11122\relax
\mciteBstWouldAddEndPuncttrue
\mciteSetBstMidEndSepPunct{\mcitedefaultmidpunct}
{\mcitedefaultendpunct}{\mcitedefaultseppunct}\relax
\EndOfBibitem
\bibitem[Atkins and Friedman(2010)Atkins, and Friedman]{Atkins2010_Book}
Atkins,~P.~W.; Friedman,~R.~S. \emph{Molecular Quantum Mechanics}, 5th ed.;
  Oxford University Press, 2010\relax
\mciteBstWouldAddEndPuncttrue
\mciteSetBstMidEndSepPunct{\mcitedefaultmidpunct}
{\mcitedefaultendpunct}{\mcitedefaultseppunct}\relax
\EndOfBibitem
\bibitem[Bernstein(2018)]{Bernstein2018_Book}
Bernstein,~D.~S. \emph{Scalar, Vector, and Matrix Mathematics: Theory, Facts,
  and Formulas}, revised and expanded ed.; Princeton University Press:
  Princeton, New Jersey, 2018\relax
\mciteBstWouldAddEndPuncttrue
\mciteSetBstMidEndSepPunct{\mcitedefaultmidpunct}
{\mcitedefaultendpunct}{\mcitedefaultseppunct}\relax
\EndOfBibitem
\bibitem[Goodman(2014)]{Goodman2014_Book}
Goodman,~F.~M. \emph{Algebra: Abstract and Concrete}, 2nd ed.; SemiSimple
  Press, 2014\relax
\mciteBstWouldAddEndPuncttrue
\mciteSetBstMidEndSepPunct{\mcitedefaultmidpunct}
{\mcitedefaultendpunct}{\mcitedefaultseppunct}\relax
\EndOfBibitem
\bibitem[Glaser \latin{et~al.}(2023)Glaser, Baiardi, Kelemen, and
  Reiher]{Glaser2023_Colibri}
Glaser,~N.; Baiardi,~A.; Kelemen,~A.~K.; Reiher,~M. {qcscine/colibri}: Release
  1.0.0. Zenodo, 2023; \url{https://doi.org/10.5281/zenodo.10276683}\relax
\mciteBstWouldAddEndPuncttrue
\mciteSetBstMidEndSepPunct{\mcitedefaultmidpunct}
{\mcitedefaultendpunct}{\mcitedefaultseppunct}\relax
\EndOfBibitem
\bibitem[Colbert and Miller(1992)Colbert, and Miller]{Colbert1992_1982}
Colbert,~D.~T.; Miller,~W.~H. A novel discrete variable representation for
  quantum mechanical reactive scattering via the S-matrix Kohn method. \emph{J.
  Chem. Phys.} \textbf{1992}, \emph{96}, 1982--1991\relax
\mciteBstWouldAddEndPuncttrue
\mciteSetBstMidEndSepPunct{\mcitedefaultmidpunct}
{\mcitedefaultendpunct}{\mcitedefaultseppunct}\relax
\EndOfBibitem
\bibitem[Light and Carrington(2000)Light, and Carrington]{Light2000_263}
Light,~J.~C.; Carrington,~T.,~Jr. In \emph{Advances in Chemical Physics};
  Prigogine,~I., Rice,~S.~A., Eds.; John Wiley \& Sons, 2000; Vol. 114; pp
  263--310\relax
\mciteBstWouldAddEndPuncttrue
\mciteSetBstMidEndSepPunct{\mcitedefaultmidpunct}
{\mcitedefaultendpunct}{\mcitedefaultseppunct}\relax
\EndOfBibitem
\bibitem[Stewart(2007)]{Stewart2007_1173}
Stewart,~J. J.~P. Optimization of parameters for semiempirical methods V:
  Modification of NDDO approximations and application to 70 elements. \emph{J.
  Mol. Model.} \textbf{2007}, \emph{13}, 1173--1213\relax
\mciteBstWouldAddEndPuncttrue
\mciteSetBstMidEndSepPunct{\mcitedefaultmidpunct}
{\mcitedefaultendpunct}{\mcitedefaultseppunct}\relax
\EndOfBibitem
\bibitem[Husch \latin{et~al.}(2018)Husch, Vaucher, and
  Reiher]{Husch2018_e25799}
Husch,~T.; Vaucher,~A.~C.; Reiher,~M. Semiempirical molecular orbital models
  based on the neglect of diatomic differential overlap approximation.
  \emph{Int. J. Quantum Chem.} \textbf{2018}, \emph{118}, e25799\relax
\mciteBstWouldAddEndPuncttrue
\mciteSetBstMidEndSepPunct{\mcitedefaultmidpunct}
{\mcitedefaultendpunct}{\mcitedefaultseppunct}\relax
\EndOfBibitem
\bibitem[Bosia \latin{et~al.}(2023)Bosia, Zheng, Vaucher, Weymuth, Dral, and
  Reiher]{Bosia2023_054118}
Bosia,~F.; Zheng,~P.; Vaucher,~A.; Weymuth,~T.; Dral,~P.~O.; Reiher,~M.
  Ultra-fast semi-empirical quantum chemistry for high-throughput computational
  campaigns with Sparrow. \emph{J. Chem. Phys.} \textbf{2023}, \emph{158},
  054118\relax
\mciteBstWouldAddEndPuncttrue
\mciteSetBstMidEndSepPunct{\mcitedefaultmidpunct}
{\mcitedefaultendpunct}{\mcitedefaultseppunct}\relax
\EndOfBibitem
\bibitem[Kamath \latin{et~al.}(2018)Kamath, Vargas-Hern{\'a}ndez, Krems,
  Carrington, and Manzhos]{Kamath2018_241702}
Kamath,~A.; Vargas-Hern{\'a}ndez,~R.~A.; Krems,~R.~V.; Carrington,~T.,~Jr.;
  Manzhos,~S. Neural networks vs Gaussian process regression for representing
  potential energy surfaces: A comparative study of fit quality and vibrational
  spectrum accuracy. \emph{J. Chem. Phys.} \textbf{2018}, \emph{148},
  241702\relax
\mciteBstWouldAddEndPuncttrue
\mciteSetBstMidEndSepPunct{\mcitedefaultmidpunct}
{\mcitedefaultendpunct}{\mcitedefaultseppunct}\relax
\EndOfBibitem
\bibitem[Ziegler and Rauhut(2019)Ziegler, and Rauhut]{Ziegler2019_4187}
Ziegler,~B.; Rauhut,~G. Localized Normal Coordinates in Accurate Vibrational
  Structure Calculations: Benchmarks for Small Molecules. \emph{J. Chem. Theory
  Comput.} \textbf{2019}, \emph{15}, 4187--4196\relax
\mciteBstWouldAddEndPuncttrue
\mciteSetBstMidEndSepPunct{\mcitedefaultmidpunct}
{\mcitedefaultendpunct}{\mcitedefaultseppunct}\relax
\EndOfBibitem
\end{mcitethebibliography}
\providecommand{\latin}[1]{#1}
\makeatletter
\providecommand{\doi}
  {\begingroup\let\do\@makeother\dospecials
  \catcode`\{=1 \catcode`\}=2 \doi@aux}
\providecommand{\doi@aux}[1]{\endgroup\texttt{#1}}
\makeatother
\providecommand*\mcitethebibliography{\thebibliography}
\csname @ifundefined\endcsname{endmcitethebibliography}
  {\let\endmcitethebibliography\endthebibliography}{}

\end{document}